\documentstyle[12pt]{article}

\begin{document}
\newcommand{\be}{\begin{equation}}
\newcommand{\ee}{\end{equation}}
\newcommand{\bee}{\begin{ eqnarray }}
\newcommand{\eee}{\end{ eqnarray }}
\newcommand{\nn}{\nonumber \\ }


\title{ Geodesics deviation equation \\ approach to chaos.}

\author{
{\sc  J. Szcz\c{e}sny and T.Dobrowolski}
\\ Institute of Physics and Informatics WSP, 
\\ Podchor\c{a}\.zych  2, 30-084 Cracow, Poland
}

\date{}
\maketitle


\begin{abstract}
Geodesics deviation equation (GDE) is introduced.
In "adiabatic" approximation exact solution  of the GDE is found.
Perturbation theory in general case is formulated.
Geometrical criterion of local instability which may
lead to chaos is formulated. 

\end{abstract}

\newpage

\section{ Introduction}

Even though classical mechanics is an old subject,
the surprising fact is that the mechanisms affecting 
its dynamical evolution have only recently been 
qualitatively understood. It is known that chaos
in classical mechanical systems emerges when 
overlapping of resonance zones occurs.
The essential problem in study of chaos lies in formulation of a 
simple mathematical criterion of sensitive dependence
on initial conditions. 

An example of analytical criterion is proposed by
Toda, Duff and Brumer \cite{Toda} local criterion of the 
transition from the ordered motion to the chaotic one.
The main idea of their approach lies in replacement 
studies of the behaviour of trajectories in phase
space which remain nearby to the selected trajectory
by studies of the behaviour of nearby trajectories in 
vicinity of selected points of the phase space.

The other local criterion of chaos is studied in this
article. It is based on observation that Euler - Lagrange
equations of motion can be rewritten as the geodesic equation
with respect to the Jacobi metric and then on measuring
the local tendency of geodesics to converge or to diverge.
The idea of this method has been originated in Krylov paper \cite{Krylov}.
The use of the deviation equation presupposes that geodesics
form a congruence which is true locally in admissible for 
trajectories region of the configuration space.
The property of being a K-system is a global property
and therefore deviation equation which is purely
local criterion needs some extension so as to obtain
a more complete (a kind of Lapunov exponent and study of 
the influence of the boundary $\partial {\cal D}_E$ on
global behaviour of the system) characteristic
of the system.

The overall plan of this article is as follows.
In section 2, we introduce in a new, simplified way
the concept of the geodesics deviation and find a
simple form of geodesics deviation equation.
Section 3 is devoted entirely to searching
for solutions of the GDE and to formulation of
the perturbation  expansion.
The last section (4) contains examples and 
comments.

\section{Introduction to geodesics deviation equation}

\subsection{Dynamics of the system in arbitrary coordinates.}
Let us consider classical mechanical system with $N$ degrees of 
freedom ($i,j = 1,2 ...,N$) described by the  lagrangian
\be 
{\cal{L}}(q,\dot{q}) = {1 \over 2} g_{ij} \dot{q}^i \dot{q}^j - V(q) ,
\ee
where $g_{ij} dq^i \otimes dq^j$ is a Riemannian metric on the configuration 
space $\cal{M}$ of the system. 
Let us assume  that the dynamics of this system could be 
equivalently described in terms of a momentum - position phase 
space by the Hamiltonian
\be 
{\cal{H}}(q,p) = {1 \over 2} g^{ij}(q) p_i p_j + V(q),
\ee
$p_i = g_{ij} \dot{q}$ denotes momentum conjugate with $i-th$ coordinate $q^i$,
and "$^. = {d \over dt}$" differentiation  with respect to time.
Dynamics of the system is governed by the Hamilton equations of motion:
\be
\dot{p_i} = - {1 \over 2} 
{\partial g^{jk} \over \partial {q}^i} p_j p_k - 
{\partial V \over \partial q^i} \mbox{~~~ and ~~~ } \dot{q}^i =  g^{ik} p_k .
\ee 
By eliminating $p_i$ variables and using identity 
${\partial g^{ik} \over \partial q^l} g_{jk} 
=  - g^{ik}{\partial g_{jk} \over \partial q^l}$ we can rewrite a system 
of the first order equations as a system of the second order equations
\be
{\ddot{q}}^i + {\Gamma^i}_{kl} \dot{q}^k \dot{q}^l = - g^{ik} \partial_k V ,
\ee 
$\partial_k = {\partial \over \partial q^k}$ and 
$  {\Gamma^i}_{kl} $ is Levi-Civita connection for metric $g_{ij}$.

\subsection{Dynamics on the boundary of an admissible region}

An  important characteristic of the mechanical system  is the shape 
of region of the configuration space
which is admissible for the physical trajectories. 
It is known that even small deformation 
of this region could lead  from deterministic to chaotic behavior 
of the  system. Well known example of such behavior is the stadium 
\cite{Strelcyn}. 
We will pay some attention to properties of this region. 

It is obvious that the total energy of the system is an integral of motion. 
Therefore for fixed (by the choice of the initial conditions) energy of 
the system any trajectory in phase space is confined to the hypersurface 
${1 \over 2} g^{ij}(q) p_i p_j + V(q) = E$. The kinetic energy of the system 
${1 \over 2} g^{ij} p_i p_j$ is positive and therefore (for fixed total energy 
$E$) a projection of the constant energy hypersurface on configuration 
space provides an  admissible for trajectories (movement) region of the configuration 
space ${\cal{D}}_E = \{q\in {\cal{M}}: V(q ) \leq E\}$. ${\cal{D}}_E$ 
depending on the energy, 
could be bounded or unbounded, connected or not. In general 
${\cal{D}}_E$ has boundary $\partial {\cal{D}}_E = \{ q \in {\cal{M}}: 
V(q) = E \}$. 
If potential $V(q)$ has no critical points on the boundary
then $\partial {\cal{D}}_E$ is $N-1$ dimensional submanifold of ${\cal{M}}$.
\vskip 0.25cm 
{\it We can easy see that}
{\bf point ${\bf q_0 \in \partial {\cal{D}}_E}$ 
in which trajectory reaches the boundary
is an isolating point in ${\bf\partial {\cal{D}}_E}$.}\hfill\break
\vskip -0.4cm
If trajectory reaches $q_0 \in \partial {\cal{D}}_E$ then speed $v^i$ in this 
point is equal to zero. It is consequence of equality 
$ {1 \over 2} v^i v^j g_{ij} + V(q_0) = E = V(q_0) $. We see that in 
$ \partial {\cal{D}}_E$ exist neighbourhood of  $q_0 \in \partial {\cal{D}}_E$ 
which does not contain any point of the trajectory. 
It means that movement along boundary $ \partial {\cal{D}}_E$ is impossible,
and that the point $q_0$ is isolating point.
\vskip 0.25cm
{\it In fact, it is easy to show that }
{\bf if potential is smooth function of \break space coordinates
then trajectories approach to ${\bf q_0 \in \partial {\cal{D}}_E}$ or depart  
from ${\bf q_0 \in \partial {\cal{D}}_E}$ perpendicularly to the boundary 
${\bf \partial {\cal{D}}_E}$.  }\hfill\break
\vskip -0.4cm
i) Any vector $\xi^i$ tangent to $\partial {\cal{D}}_E$ satisfy equation 
$\partial_i V(q_0) \xi^i = 0$ which could be rewritten 
in the form $g_{ij} {(grad V)^i}\mid_{q=q_0}  \xi^j = 0$. 
This is just orthogonality condition in the sense of scalar product 
defined by the matrix $g_{ij}(q_0)$.

ii) On the other hand if we assume that the trajectory reaches the point 
$q_0$ in instant of time $t_0$ then for $t$ close to $t_0$ we have 
$q^i(t) = q^i_0 + \dot{q}^i(t_0) (t - t_0) + 
{1 \over 2} \ddot{q}^i(t_0) (t - t_0)^2 + O((t-t_0)^2)$. 
Reminding that $v^i = \dot{q}^i(t_0) = 0$ and using equations of motion 
(4)  we obtain $q^i(t) = q^i_0 - 
{1 \over 2} g^{ij}(q_0) \partial_j V(q_0) (t - t_0)^2 + O((t-t_0)^2)$. 
After differentiation of the last formula with respect to $t$ we evaluate 
velocity in neighbourhood 
of the point $q_0$, ~ $\dot{q}^i(t) =  (grad V)^i\mid_{q=q_0} (t_0 - t)$.

From i) and ii) it follows  that trajectories are orthogonal to the 
boundary $\partial {\cal{D}}_E$.  

The stadium (which is not smooth) does not satisfy
assumption of the above statement.

\subsection{ Geometric form of the equations of motion }

Let us turn to dynamics of the system in the interior of the admissible 
region $Int {\cal{D}}_E = \{q \in {\cal{M}}: V(q) < E   \}$.
\vskip 0.25cm
{\it Crucial to our future construction  is an observation that}
{\bf geodesic equation in the Riemannian geometry defined by the Jacobi metric} 
${\bf \hat{g}^{ij} =  \psi g^{ij}}$ { \it (with $\psi(q) = 2(E - V)$ and natural parameter s 
so that ${d s \over d t} = 2 (E - V)$)}
{\bf on the set ${\bf Int {\cal{D}}_E}$ 
is equivalent to the equations of motion (4). } \hfill\break
\vskip -0.4cm
Geodesic equation for metric $\hat{g}_{ij} = \psi(q) g_{ij}$ has the well known form
\be
{d^2 q^i \over d s^2} + 
{\hat{\Gamma}^i}_{jk} {d q^j \over d s} {d q^k \over d s} = 0 ,
\ee
where $s$  is a  natural parameter in the sense of  the Jacobi metric 
$\left(   \hat{g}_{ij}(q(s)) {d q^i \over d s} {d q^j \over d s} = 1 \right)$ 
and ${\hat{\Gamma}^i}_{jk}$ 
is a Christoffel symbol with respect to the same 
metric $\hat{g}_{ij}$. 
If we express Christoffel symbols of the Jacobi metric 
by  Christoffel symbols 
${{\Gamma}^i}_{jk}$ of the metric $g_{ij}$
\be
{\hat{\Gamma}^i}_{jk} = {\Gamma^i}_{jk} + {1 \over 2} [ 
\partial_j(ln\psi) {\delta^i}_k  + \partial_k(ln\psi) {\delta^i}_j - \partial_r(ln\psi) 
g^{ri} g_{jk}] ,
\ee
and exchange the natural parameter $s$ for time  $t$ 
then we will obtain the rearranged  equation (5) 
\be
{d^2 q^i \over d t^2} + 
{{\Gamma}^i}_{jk} {d q^j \over d t} {d q^k \over d t} = 
\left[ \left( {d s \over d t}\right)^2 {d^2 t \over d s^2} + {d q^i \over d t} \partial_j ln \psi \right] {d q^i \over d t} + {1 \over 2 \psi} \left({d s \over d t} \right)^2 g^{ik} \partial_k ln \psi .
\ee
Now we would like to chose $\psi$ and parameter $s=s(t)$ so as to geodesic equation (5) be equivalent to equation (4). Let us impose the simplest  conditions
which eliminate unwanted terms from equation (7)
\be
\left( {d s \over d t}\right)^2 {d^2 t \over d s^2} + {d q^i \over d t} \partial_j 
ln \psi = 0 ,
\ee
\be
{1 \over 2 \psi} \left({d s \over d t} \right)^2 g^{ik} \partial_k ln \psi = - \partial_j V .
\ee 
Making further manipulations on parameter derivatives 
$\left({d^2 t \over d s^2} =  {d t \over d s} {d \over d t} \left(  {1 \over {d s \over dt}}\right) = 
- {1 \over \left( {d s \over d t }\right)^3} {d^2 s \over d t^2}\right)$,
we can simplify the first constrain (8) to the form 
\be
{d u \over d t} = u {d ln \psi \over d t} ,
\ee
where $u(t) = \left( {d s \over d t} \right)^2$.

The solution of this equation 
$\left({d s \over d t}\right)^2 = u(t) = A \psi^2$
married with condition (9) gives
${1 \over 2} \partial_l \psi = - \partial_l V$, which
provides the final answer
$$\psi = 2 (M - V) .$$
$M$ is an integration constant ($A$ equal one was taken for simplicity).

The only unknown is interpretation of an arbitrary constant.
Reminding that $s$ is a natural parameter for Jacobi Metric
$$
1 = \hat{g}_{ij} {d q^i \over d s} {d q^j \over d s} =
2 (M - V) g_{ij} \left({d t \over d s} \right)^2 {d q^i \over d t} {d q^j \over d t} =
{1 \over 2(M - V)} g_{ij} {d q^i \over d t} {d q^j \over d t}, 
$$
we find that $M$ is the total energy of the system
$$
M = {1 \over 2} g_{ij} {d q^i \over d t} {d q^j \over d t} + V = E .
$$
Therefore  
\be
\psi = 2 (E - V) .
\ee

{\it We can also check that} ${\bf s}$ {\bf is a good evolution 
parameter in the region} ${\bf Int{\cal{D}}_E}$ { \it i.e.
$t \rightarrow s(t)$ is monotonically increasing
function of time.} \hfill\break
\vskip -0.4cm
In admissible for trajectories region of configuration space
${d s \over d t}(t) = 2 (E - V) \geq 0$, 
$
{d^2 s \over d t^2}(t) = - 2 \dot{q}^i \partial_i V
$,
and
$ {d^3 s \over d t^3}(t) = 
- 2 \ddot{q}^i \partial_i V
- 2 \dot{q}^i \dot{q}^j \partial_i \partial_j V$.
Let us fix the origin of the  parameter transformation
$s(t=0) = 0$.
If for some value of the parameters 
$s(t_0) = s_0$
the system reaches the boundary of the admissible region
$q_0 \in \partial {\cal{D}}_E$ then 
$$
{d s \over d t}(t_0) = {d^2 s \over d  t^2}(t_0) = 0 , ~~~ 
{d^3 s \over d t^3} = 2 g^{ij}(q_0) \partial_i V(q_0) \partial_j V(q_0) > 0 .
$$
In the last inequality we used equations of motion (4).
We see that 
$s_0 = s(t_0)$ is 
an inflection point of the considered function
$t \rightarrow s(t)$. In the region $Int{\cal{D}}_E$
$s(t)$ is monotonically increasing function of time.

\subsection{Origin of the GDE}
In the previous section we discovered that Euler - Lagrange equations (4)
could be transformed to the geodesic equation 
\be
\hat{\nabla}_u u = 0 ,
\ee
where $u$ is the tangent vector to the geodesic and $\hat{\nabla}$ is the 
covariant derivative with respect to the Jacobi metric. 
Our purpose is to investigate the relative motion of geodesics
in a given domain $Int {\cal{D}}_E$ of the configuration space.
We choose  a transverse curve $C$ with respect to the congruence
of geodesics, i.e., the curve which only once crosses each of the geodesics
belonging to the congruence. We assume that the point at which the
curve $C$ crosses a given geodesic is the zero point of the parameter $s$ along 
this geodesic. Now, we find the curve $C_s$ which is the copy of the curve $C$.
We construct $C_s$ by transporting each point of the curve $C$, 
along the geodesic to the point on this geodesic at which the 
value of the parameter is $s$. In this way, in the domain in which the congruence 
of geodesics is determined, we obtain a 2-dimentional surface. The tangent vector fields to this surface at the point for which the values of parameters are
$(\lambda, s)$ will be denoted by
$ u(\lambda, s) = {\partial \over \partial s} \mid_{(\lambda, s)} $,
and
$ \xi(\lambda, s) = {\partial \over \partial \lambda} \mid_{(\lambda, s)} $.
The field $u$ is also tangent to the geodesic. By construction of the surface 
the Lie bracket of the vector fields  vanish $[u,\xi] = 0$.
Since the connection $\hat{\Gamma}$ is torsionfree, one has
\be
\hat{T}(u,\xi) = \hat{\nabla}_u \xi - \hat{\nabla}_{\xi} u - [u,\xi] = 0,
\ee
and therefore
\be 
\hat{\nabla}_u \xi = \hat{\nabla}_{\xi} u .
\ee
Now we would like to find acceleration of the changes of the vector field $\xi$
along a given geodesic. From the definition of the curvature one could obtain
\be
\hat{R}(u,\xi) u = \left(
\hat{\nabla}_u \hat{\nabla}_{\xi} -
\hat{\nabla}_{\xi} \hat{\nabla}_u -
\hat{\nabla}_{[u,\xi]}
\right) u = 
\hat{\nabla}_u \hat{\nabla}_{\xi} u .
\ee
With the help of equations (14-15) we can find  acceleration of the $\xi$ field
\be
\hat{\nabla}_u \hat{\nabla}_u \xi = \hat{R}(u,\xi) u.
\ee
The vector field $\xi$ informs us about the relative position of points on 
neighbouring geodesics which start from points on  arbitrarily chosen
curve $C$. We are interested only in the relative motion with respect
to geodesics and not in the motion of points along these geodesics. The 
component of  $\xi$ parallel to the vector  $u$ contains useless information.
Therefore, we will eliminate this component focusing only on orthogonal to $u$
component of $\xi$. Let us decompose 
$\xi = n + \lambda u$, 
where
$\lambda \in I\!\!R $
and
$\hat{g}(n,u) = 0$.
We also assume unit normalization of the tangent field
$\hat{g}(u,u) = 1$.
Under above assumption the parameter $\lambda$ is just a projection 
$u$ on $\xi$,  
$\hat{g}(\xi,u) = \hat{g}(n + \lambda u,u) = \lambda \hat{g}(u,u) = \lambda$.
The vector field $n$ is called geodesics deviation 
$n = \xi - \hat{g}(\xi,u) u$.
Since $\xi = n + \hat{g}(\xi,u) u$ both sides of the 
equation (16) could be transformed to the form which is free of 
parallel (to the vector $u$) component of $\xi$
\be
\hat{\nabla}_u \hat{\nabla}_u \xi = 
\hat{\nabla}_u \left(
\hat{\nabla}_u n + \hat{g}(\hat{\nabla}_u \xi, u) u
\right) =
\hat{\nabla}_u \left(
\hat{\nabla}_u n + \hat{g}(\hat{\nabla}_{\xi} u, u) u
\right) =
\hat{\nabla}_u \hat{\nabla}_u n ,
\ee
\be
\hat{R}(u,\xi) u = \hat{R}(u,n) u + \hat{g}(\xi,u) \hat{R}(u,u) u =
\hat{R}(u,n) u .
\ee
As a result of (17-18) equation (16) can be rewritten 
in the form known as a geodesics deviation equation
\be
\hat{\nabla}_u \hat{\nabla}_u n
 = \hat{R}(u,n) u .
\ee
Equation (19) answers the question with which acceleration 
field $n$ changes along a given geodesic.
It should be stressed that equation (19) measures {\bf local}
tendency of geodesics to converge or to diverge and it works
if the vector $n$ is small. It also works in a neighbourhood
of geodesics which are parallel to each other in some
interval. 

\subsection{Newtonian form of the GDE}

Let us rewrite the geodesics deviation equation in the form
similar to the Newtonian equation of motion. This rearrangement
allows us to use the known methods of solving equations
of motion.

To this  end we shall use the Riemann tensor 
$\hat{R}(A,B,C,D) \equiv \hat{g}(A,\hat{R}(C,D)B)$ where
$A,B,C,D$ are vector fields on configuration space ${\cal{M}}$.
The components of this tensor are
$\hat{R}_{ijkl} = \hat{R}\left( 
{\partial \over \partial q^i},{\partial \over \partial q^j},
{\partial \over \partial q^k},{\partial \over \partial q^l} \right)
= \hat{g}_{ir} {\hat{R}^r}_{jkl}$.
Using this tensor we can rearrange the right hand side
of the geodesics deviation equation (19)
\be 
[\hat{R}(u,n) u]^i = 
- {1 \over 2} \hat{g}^{ij} \hat{R}_{klrm} (u^k {\delta^l}_j u^r n^m 
+ u^k n^l u^r {\delta^m}_j) = - {1 \over 2} \hat{g}^{ij} 
{\partial \over \partial n^j} [\hat{R}(u,n,u,n)] .
\ee
Now we introduce the new gradient operator 
$[grad_n]^i = \hat{g}^{ij} {\partial \over \partial n_j}$
and define a "potential"
${\cal{V}}_u(n) = {1 \over 2} \hat{R}(u,n,u,n)$.
This leads to the equation
\be
{D^2 n \over d s^2} \equiv \hat{\nabla}_u \hat{\nabla}_u n = - grad_n {\cal{V}}_u(n) 
\ee
which looks like Newton equation of motion. 

Newtonian nature of equation (21) is even more transparent
in the Fermi frame 
$(E_1, E_2, ..., E_{N-1}, E_N)$,
defined by $\hat{\nabla}_u E_a = 0$
and
$\hat{g}(E_a,E_b) = \delta_{ab}$
where
$a,b = 1,2, ... ,N-1,N $.

If we chose the last vector of this frame as a tangent
to the geodesic
$E_N = u$
and denote
$\alpha, \beta = 1,2, ... ,N-1$
then deviation equation (21) takes the form
\be
{d^2 n^{\alpha} \over d s^2} = - {\partial \over \partial n^{\alpha}} {\cal{V}}_u(n) .
\ee
The function ${\cal{V}}$ 
is the quadratic form  
with respect
to the $n^{\alpha}$ variables 
${\cal{V}}_u(n) = A_{\alpha \beta} n^{\alpha} n^{\beta}$,
where $[A_{\alpha \beta}]$
is $(N-1)\times(N-1)$ matrix defined by the proper components
of the Riemann tensor in the Fermi base 
$A_{\alpha \beta} = \hat{R}_{\alpha N \beta N}$.
Equation (22) describes the system of $N-1$
"coupling  oscillators" with "time"-s 
dependent frequencies.
If we consider the relative motion of geodesics only locally
(in the neighbourhood of the point $q_*=q(s_*)$) then 
matrix  $[A_{\alpha \beta}]$ is constant and values of its
components are defined by the values of the Riemann tensor
in the point $q_*$. It is clear that if ${\cal{V}}$ is
positively defined then motion of geodesics in the
neighbourhood of the considered point is stable. In this case geodesics
approach each other. If ${\cal{V}}$ is undefined or 
negatively defined then geodesics diverge i.e. motion is
unstable \cite{Arnold} and we have sensitive dependence
on initial conditions.

Above analysis 
has purely local character and does not take into
account the influence of the boundary $\partial {\cal{D}}_E$
on the geodesics motion.
It should be stressed that it does not determine
existence of {\bf global} chaos in the system.
Anyhow, when the admissible region ${\cal{D}}_E$ has no boundary
then with the use of (22) new criterion can be
formulated which determines the behaviour of 
the system completely \cite{Anosov}.

\subsection{GDE in terms of an external curvature}

The potential ${\cal{V}}$ has a nice geometric interpretation.
At a given point of the manifold, for any non-collinear vector fields 
$A$ and $B$ one can define the sectional
curvature $\hat{K}_{AB}$ in the two-direction determined by these
fields in the following way
$\hat{R}(A,B,A,B) = 
\hat{K}_{AB} [ \hat{g}(A,A) \hat{g}(B,B) - \hat{g}(A,B) \hat{g}(A,B)]$.
Therefore, reminding that $\hat{g}(n,u) = 0$ and $\hat{g}(u,u) = 1$ we have
\be
{\cal{V}}_u(n) = {1 \over 2} \hat{R}(u,n,u,n) = {1 \over 2} \hat{K}_{un} \hat{g}(n,n) .
\ee
Hence, the potential is proportional to the sectional curvature
$\hat{K}_{un}$ and square of the deviation field.

Let us confine to the two dimensional case
where $(E_1,E_2)$ is Fermi frame.
If we chose $E_2 = u$
and
$n = x E_1$
then
${\cal{V}}_u(n) = {1 \over 2} \hat{R}(u,n,u,n)
= {1 \over 2} \hat{R}_{2121} x^2 $.
In two dimensions the sectional curvature has only one independent
component which in Fermi frame is equal to the Gauss curvature
$\hat{K} = \hat{R}_{1212}$.
The deviation equation takes now the simple form
\be
{d^2 x \over d s^2} = - \hat{K}(s) x .
\ee
It is transparent that when the Gauss curvature is positive then geodesics approaches
each other. In opposite case ($\hat{K}<0$) they diverge.

\section{Solutions of the geodesics deviation equation}

\subsection{Adiabatic approximation}

Geodesics deviation equation in two dimensions can be rewritten as a system 
of first order equations
\be
\dot{X} = {\cal{A}}(s) X
\ee
where
$" ^. " = {d \over d s}$.
In the case of two dimensional system the vector of variables has two
components
$X = \left[\begin{array}{c} x \\ \dot{x}
                                    \end{array} \right]$,
and the matrix ${\cal{A}}$ has the form 
${\cal{A}} = \left[\begin{array}{cc} 0 & 1 \\  - \hat{K}(s) & 0
                                   \end{array} \right] $ \hfill\break
{\bf i)} At the beginning let us consider 
{\bf positive curvature case}
$ \hat{K}(s) > 0$. \hfill\break
Having non-singular matrix
${\cal{P}}_1 = \left[\begin{array}{cc} 1 & 0 \\ 0 & - \sqrt{\hat{K}(s)}
                                   \end{array} \right] $
we can transform vector $X$,
$X(s) = {\cal{P}}_1(s) Y(s)$ 
and the whole equation (25) to the form
\be
\dot{Y} = [ {{\cal{P}}_1}^{-1} {\cal{A}} {\cal{P}}_1 - 
{{\cal{P}}_1}^{-1} \dot{{\cal{P}}}_1] Y .
\ee
During this transformation only locality of the matrix ${\cal{P}}_1$
needs some care.
Explicit manipulations on matrices ${\cal{A}}$ and ${\cal{P}}_1$
lead to the equation
\be
\dot{Y} =  D_1
\left(I - \left[\begin{array}{cc} 0 & { \dot{\hat{K}} \over 2 \hat{K}^{3 \over 2}} 
                          \\ 0 & 0  \end{array} \right]   \right) Y ,
\ee
where $I$ is identity matrix and
$D_1 = {{\cal{P}}_1}^{-1} {\cal{A}} {\cal{P}}_1 = \sqrt{\hat{K}} \left[\begin{array}{cc} 0 & -1 \\ 1 & 0
                                   \end{array} \right] . $

We see from (27) and the form of $D_1$ that $\sqrt{\hat{K}}$ has the
same dimension as "s" derivative of $\hat{K}$. This observation allows us to
use a ratio $\dot{\hat{K}} \over \hat{K}^{3 \over 2}$
as a dimensionless perturbation parameter.

We assume that in the neighbourhood of the considered point
of the admissible region Gauss curvature changes very slowly in
comparison to its value i.e. $\mid \dot{\hat{K}} \mid \ll \hat{K}^{3 \over 2}$.
Next we keep only zero order term in equation (27) 
\be
\dot{Y} =  D_1(s) Y .
\ee
We call this simplification {\bf adiabatic} approximation.
Note that $D_1$ is still curvature $ \hat{K}$ dependent.
To make further progress we denote components of the 
vector $Y$ in the following way
$Y = \left[\begin{array}{c} y \\  \tilde{y}
                                    \end{array} \right] , $ 
then equation (28) appears to be a system of coupled first order equations
\be
\dot{y}(s) = - \sqrt{\hat{K}(s)} \tilde{y}(s) , ~~~~~
\dot{\tilde{y}}(s) = \sqrt{\hat{K}(s)} y(s) .
\ee
Introducing a complex variable $z(s) = y(s) + i \tilde{y}(s)$
we rearrange the system (29)
\be
\dot{z}(s) = i \sqrt{\hat{K}} z(s) ,
\ee
so as to easy find solution
\be
z(s) = z(0) exp\left[i {\int_0}^s d s' \sqrt{\hat{K}(s')} \right] .
\ee
Coming back to the real vector $Y$ we have
\be
Y(s) = \left[ \begin{array}{cc} 
cos\left({\int_0}^s d s' \sqrt{\hat{K}(s')} \right) &
- sin\left({\int_0}^s d s' \sqrt{\hat{K}(s')} \right) \\
sin\left({\int_0}^s d s' \sqrt{\hat{K}(s')} \right) &
cos\left({\int_0}^s d s' \sqrt{\hat{K}(s')} \right) \end{array} \right] Y(0) .
\ee
Transforming (32) by the transformation 
$ X(s) = {\cal{P}}_1(s) Y(s) $
and then initial values of $Y(0)$ by the transformation
$Y(0) = {{\cal{P}}_1}^{-1}(0) X(0) $
we can express the vector $X(s)$ by its initial values
\be
X(s) = \left[ \begin{array}{cc} 
cos\left({\int_0}^s d s' \sqrt{\hat{K}(s')} \right) &
{1 \over \sqrt{\hat{K}(0)}} sin\left({\int_0}^s d s' \sqrt{\hat{K}(s')} \right) \\
- \sqrt{\hat{K}(s)} sin\left({\int_0}^s d s' \sqrt{\hat{K}(s')} \right) &
\sqrt{\hat{K}(s) \over \hat{K}(0)} cos\left({\int_0}^s d s' \sqrt{\hat{K}(s')} \right) \end{array} \right] X(0) .
\ee
Therefore adiabatic problem in $ \hat{K} > 0$ case has the solution
\be
x(s) =  
x(0) cos\left({\int_0}^s d s' \sqrt{\hat{K}(s')} \right) +
{\dot{x}(0) \over \sqrt{\hat{K}(0)}} 
sin\left({\int_0}^s d s' \sqrt{\hat{K}(s')} \right) .
\ee
We see that equation (34) describes stable relative oscillations
of geodesics. The frequency of these oscillations
changes with "time" - s.
\hfill\break
{\bf ii)} solution of the {\bf negative curvature case} $ \hat{K} < 0$ 
\hfill\break

Construction of the solution for negative Gauss curvature is analogous
to the proceeded construction of the positive curvature solution.
The only difference lies in the form of the transformation matrix
${\cal{P}}_2 = \left[\begin{array}{cc} 1 & 1 \\ - \sqrt{\mid \hat{K} \mid} & \sqrt{\mid \hat{K} \mid}
                                   \end{array} \right] . $
After this transformation the deviation equation takes
similar to the equation (26) form
\be
\dot{Y} = [ {{\cal{P}}_2}^{-1} {\cal{A}} {\cal{P}}_2 - 
{{\cal{P}}_2}^{-1} \dot{{\cal{P}}_2}] Y ,
\ee
where
${\cal{A}} = \left[\begin{array}{cc} 0 & 1 \\ \mid \hat{K}(s) \mid & 0 
                                   \end{array} \right] . $
Detailed structure of equation (35) differs significantly from 
equation (27)
\be
\dot{Y} =  D_2(s)
\left(I - 
{ \dot{\mid \hat{K} \mid} \over 4 \mid \hat{K} \mid^{3 \over 2}} 
\left[\begin{array}{cc} -1 &  1
                          \\ -1 & 1  \end{array} \right]   \right) Y ,
\ee
where
$D_2 = {{\cal{P}}_2}^{-1} {\cal{A}} {\cal{P}}_2 = 
\sqrt{\mid \hat{K} \mid} \left[\begin{array}{cc} -1 & 0 \\ 0 & 1
                                   \end{array} \right] . $

For slowly varying curvature i.e. 
$\mid \dot{\hat{K}} \mid \ll \mid\hat{K}\mid^{3 \over 2}$,
we can approximate equation (36) by 
\be
\dot{Y} =  D_2(s) Y .
\ee
Fortunately equation (37) describes a system of decoupled first 
order equations 
\be
\dot{y}(s) = - \sqrt{\mid \hat{K}(s) \mid} y(s), ~~~~~
\dot{\tilde{y}}(s) = \sqrt{\mid \hat{K}(s) \mid} \tilde{y}(s),
\ee
which has a simple solution
\be
y(s) = y(0) exp\left[- {\int_0}^s d s' \sqrt{\mid \hat{K}(s') \mid} \right], ~~
\tilde{y}(s) = \tilde{y}(0) exp\left[{\int_0}^s d s' \sqrt{\mid \hat{K}(s') \mid} \right] .
\ee
Coming back to the $X$ vector we have the final answer
\be
X(s) = \left[ \begin{array}{cc} 
cosh\left({\int_0}^s d s' \sqrt{\mid\hat{K}(s')\mid} \right) &
{1 \over \sqrt{\mid \hat{K}(0)} \mid} sinh\left({\int_0}^s d s' \sqrt{\mid \hat{K}(s') \mid} \right) \\
\sqrt{\mid \hat{K}(s) \mid} sinh\left({\int_0}^s d s' \sqrt{\mid \hat{K}(s') \mid} \right) &
\sqrt{\mid \hat{K}(s) \mid \over \mid \hat{K}(0) \mid} cosh\left({\int_0}^s d s' \sqrt{\mid \hat{K}(s') \mid} \right) \end{array} \right] X(0) .
\ee
First component of this vector
\be
x(s) =  
x(0) cosh\left({\int_0}^s d s' \sqrt{\mid \hat{K}(s')\mid} \right) +
{\dot{x}(0) \over \sqrt{\mid \hat{K}(0)\mid}} sinh\left({\int_0}^s d s' \sqrt{\mid \hat{K}(s')\mid} \right) 
\ee
describes exponential divergence of the geodesics with "time"-s.

\subsection{Exact solutions of the GDE}

{\bf i)} Let us come back to the geodesics deviation equation (27)
in the $\hat{K}>0$ regime 
\be
\dot{Y}(s) = [D_1(s) + \varepsilon_1(s) B_1(s)] Y(s) ,
\ee
where
$
B_1(s) = \left[ \begin{array}{cc} 
0 & 0 \\
0 & - {1 \over 2} \sqrt{\hat{K}(s)} \end{array} \right] 
$
and by 
$ \varepsilon_1(s) = { \dot{\hat{K}}(s) \over \hat{K}^{3 \over 2} }$
we denote the perturbation parameter.

To define perturbation 
procedure we introduce a kind of interaction picture.
Having exact solutions of the adiabatic problem $Y(s)$ we
define vector $U(s)$ which carry part of the evolution of the 
system which goes beyond the adiabatic approximation
solution
\be
Y(s) = \left[ \begin{array}{cc} 
cos\left({\int_0}^s d s' \sqrt{\hat{K}(s')} \right) &
- sin\left({\int_0}^s d s' \sqrt{\hat{K}(s')} \right) \\
sin\left({\int_0}^s d s' \sqrt{\hat{K}(s')} \right) &
cos\left({\int_0}^s d s' \sqrt{\hat{K}(s')} \right) \end{array} \right] U(s) .
\ee
With the use of the new vector field the equation (42) can be rewritten
in the form
\be
\dot{U}(s) = \varepsilon_1(s) C_1(s) U(s), ~~~~ U(0) = Y(0),
\ee
where 
$$
C_1(s) = - {1 \over 2} \sqrt{\hat{K}} \left[ \begin{array}{cc} 
sin^2\left({\int_0}^s d s' \sqrt{\hat{K}(s')} \right) &
{1 \over 2} sin\left( 2 {\int_0}^s d s' \sqrt{\hat{K}(s')} \right) \\
{1 \over 2} sin\left( 2 {\int_0}^s d s' \sqrt{\hat{K}(s')} \right) &
cos^2\left({\int_0}^s d s' \sqrt{\hat{K}(s')} \right) \end{array} \right] .
$$
The formal solution of the problem 
\be
U(s) = \left[ I + \sum_{k=1}^{\infty} 
\int_0^s ds_1  \int_0^{s_1} ds_2 ... \int_0^{s_{k-1}} ds_k  
\varepsilon_1(s_1) \varepsilon_1(s_2) ...\varepsilon_1(s_k)
C_1(s_1) C_1(s_2) ... C_1(s_k)
\right] Y(0)
\ee
could be treated as a definition of perturbation theory.
If we cut the series on arbitrary $k=N$ then we obtain 
approximate solution of the problem (44). The precision of cuted solution (45)
is only limited by the number of considered terms $N$. \hfill\break
{\bf ii)} In the {\bf negative Gauss curvature} regime 
geodesics deviation equation has the form
\be
\dot{Y}(s) = [D_2(s) + \varepsilon_2(s) B_2(s)] Y(s),
\ee
where
$
B_2(s) ={1 \over 4} \sqrt{\mid \hat{K} \mid} \left[ \begin{array}{cc} 
-1 & 1 \\
1 & - 1 \end{array} \right] 
$
and perturbation parameter 
$ \varepsilon_2(s) = { \dot{\mid \hat{K} \mid}(s) \over {\mid \hat{K} \mid}^{3 \over 2} } . $
Now we transfer a part of the time dependence of the $Y(s)$ on vector $U(s)$
which carries non-adiabatic part of the evolution 
\be
Y(s) = \left[ \begin{array}{cc} 
exp\left(- {\int_0}^s d s' \sqrt{\mid \hat{K}(s') \mid} \right) &
0 \\
0 &
exp\left({\int_0}^s d s' \sqrt{\mid \hat{K}(s') \mid} \right) \end{array} \right] U(s) .
\ee
Hence $U(s)$ satisfy
\be
\dot{U}(s) = \varepsilon_2(s) C_2(s) U(s), ~~~~ U(0) = Y(0),
\ee
where
$
C_2(s) = {1 \over 4} \sqrt{\mid \hat{K} \mid} 
\left[ \begin{array}{cc} 
-1 & exp\left( 2 {\int_0}^s d s' \sqrt{\mid \hat{K}(s') \mid} \right) \\
exp\left( - 2 {\int_0}^s d s' \sqrt{\mid \hat{K}(s') \mid} \right) & -1 \end{array} \right] U(s) .
$
The solution in this case
\be
U(s) = \left[ I + \sum_{k=1}^{\infty} 
\int_0^s ds_1  \int_0^{s_1} ds_2 ... \int_0^{s_{k-1}} ds_k  
\varepsilon_2(s_1) \varepsilon_2(s_2) ...\varepsilon_2(s_k)
C_2(s_1) C_2(s_2) ... C_2(s_k)
\right] Y(0) ,
\ee
could also be used to construct the perturbation expansion. 

\section{ Stability in examples.}
In further discussion we will  use formula which express 
Gauss curvature through total energy of the system, potential
and its derivatives. 
Let us find explicit relation between the Riemann curvature
tensor ${\hat{R}^i}_{ijk}$ for Jacobi metric and curvature
tensor ${R^i}_{ijk}$ for metric $g_{ij}$
\be
\hat{{R}^i}_{jkl} = {R^i}_{jkl} + {1 \over N - 2} [ 
{\delta^i}_l C_{jk} - {\delta^i}_k C_{jl}
+ g_{kj} {C^i}_l - g_{jl} {C^i}_k ] ,
\ee
where auxiliary tensor $C_{ij}$ is built of total energy and potential
in the following way
$$
C_{ij} = - {(N - 2) \over 4 (E - V)^2 } \left[ 2 (E - V) \nabla_i \nabla_j V
+ 3 (\nabla_i V) (\nabla_j V) - {1 \over 2} g_{ij} {{\cal {N}}_V}^2 \right] ,
$$
$N$ is number of dimensions and ${{\cal {N}}_V}^2$ is square of the gradient of the potential ${{\cal{N}}_V}^2 = g^{ij} (\partial_i V) (\partial_j V) $
which is positive quantity.

Equation (50) allows to express the curvature scalar for Jacobi metric
through the curvature scalar for metric $g_{ij}$
\be
\hat{R} = \hat{g}^{ij} \hat{R}_{ij} = \hat{g}^{ij} {\hat{R}^k}_{ikj} =
{1 \over 2 (E - V)} R + {(N - 1) \over 8 (E - V)^3} \left[
4 (E - V) \Delta V + (6 - N) {{\cal {N}}_V}^2 \right] ,
\ee
where  $R = g^{ij} R_{ij} = g^{ij} {R^k}_{ikj}$
and $\Delta V$ denotes  Laplacian  $\Delta V = g^{ij} \nabla_i \nabla_j V $.

Large number of mechanical systems is determinated by metric $g_{ij}$
which is just flat metric transformed to curvilinear coordinates. If we confine
our interest to this class of systems then first term in equation
(51) disappears ($R=0$). 
Let us consider two dimensional case ($N = 2$) where exists extremely simple relation between scalar curvature and  Gauss curvature of  a given  surface 
$\hat{K} = {1 \over 2} \hat{R}$.
Equation (51) in two dimensions leads to simple form of Gauss curvature 
\be
\hat{K} = 
{ 1 \over 4 (E - V)^3} \left[
 (E - V) \Delta V +  {{\cal {N}}_V}^2 \right] .
\ee
In considered region of  configuration space  $Int {\cal D}_E$ we have positive $(E - V)>0$ and the only term which can affect the overall signe
of the Gauss curvature is the Laplacian of the potential $\Delta V$.  

Now we are ready to  test criterion  (52) on two dimensional systems.

{\it Example 1} \hfill\break
Let us consider lagrangian which describes small oscillations
$$
{\cal L} = {1 \over 2} (\dot{x}^2 + \dot{y}^2) - 
(\alpha x^2 + 2 \beta x y + \gamma y^2) .
$$
Potential in this case is quadratic form of coordinates.
We can check that if  $V$ is positively defined (i.e. ~$\alpha > 0$
and $\alpha \gamma > \beta^2 $) then Laplacian of the potential
is positive $\Delta V = 2 (\alpha + \beta) > 0$ and therefore
$ \hat{K} > 0$. 
According to our criterion (52) trajectories do not diverge. 

{\it Example 2} \hfill\break
Another example is two - body problem which is defined
by the lagrangian
$$
{\cal L} = {1 \over 2} \mu (\dot{r}^2 + r^2 \dot{\varphi}^2) - V(r),
$$
where $\mu = {m_1 m_2 \over m_1 + m_2}$ is reduced mass of the
system consisting of  two masses $m_1$, $m_2$ separated by the distance $r$.
This problem is integrable and angular momentum is its integral of motion
$L = \mu r^2 \dot{\varphi}$.

Jacobi metric for this system is equal  $\hat{g}_{ij} = 2 (E - V(r)) g_{ij}$, where $[g_{ij}] = \mu \left[ \begin{array}{cc} 
1 & 0 \\ 0 & r^2 \end{array} \right] $.
Gauss curvature of the geometry fixed by Jacobi metric is
$$
\hat{K} = {1 \over 4 \mu (E - V)^3} 
\left[ (E - V) \left( V''(r) + {1 \over r} V'(r) \right) + 
\left( V'(r) \right)^2 \right] .
$$
For particular form of the potential $V(r) = - {\alpha \over r}$,
$(\alpha > 0)$ Gauss curvature is even more simple
$$
\hat{K}(r) = - { \alpha E \over 4 \mu ( \alpha + E r)^3} .
$$
In the  Kepler problem we have two cases.\hfill\break
i)  We have eliptic motion in the regime $E < 0$ . In this case
$\hat{K} > 0$ and trajectories do not diverge.\hfill\break
ii)  If total energy is positive $E > 0$ then $\hat{K} < 0$ and
we have hiperbolic motion. Although trajectories diverge the system is integrable. 

The exponential instability (i.e. sensitivity to the initial conditions)
is one of the important features of the chaotic evolution of
dynamical systems and therefore determining of the simple and possibly
elegant criterion decisive when this instability will take
place, is so important.

Our formula (22) and the comments to it, show the elegant, geometric
formulation of the criterion of the local stability of relative movement 
of geodesics.
The chaotic behaviour of dynamical systems is very complicated phenomenon
and the exponential instability 
(the negative sign of Gauss curvature $\hat{K}$) is not mostly the 
sufficient condition in order to the corresponding systems will
behave chaotically (as the example of Kepler problem shows).
It is known from Anosov paper \cite{Anosov} that
{\it on the compact manifold without boundary the geodesics family
corresponding to the geometry with everywhere negative Gauss curvature
$\hat{K}$ behaves chaotically}.
Unfortunately, space  ${\cal D}_E$ admissible for movement,
usually has non - empty boundary $\partial {\cal D}_E$,
on which the corresponding geometry, determined by Jacobi metric
$[\hat{g}_{ij}]$ becomes singular and which has important
influence on the geometry of the family of geodesics. Analysis of this
influence requires application of more advanced global methods.

\eject


\begin{thebibliography}{99}
\bibitem{Toda}
M.Toda, Phys.Lett. A {\bf 48} (1974) 5. \\ \
P.Brumer, J.W.Duff, J.Chem.Phys. {\bf 65} (1976) 3566. \\ \
\bibitem{Krylov}
N.S.Krylov, "Works on the Foundations of Statistical Physics,"
Princeton, New Jersey, 1979. \\ \
\bibitem{Strelcyn}
L.A.Bunimovich, Funct.Anal.Appl. {\bf 8} (1974) 254.\\ \
G.Benettin and J.M.Strelcyn, Phys.Rev. A {\bf 17} (1978) 773.\\ \
\bibitem{Arnold}
V.I.Arnold, "Mathematical Methods of Classical Mechanics,"
Springer Verlag, New York, 1978. \\ \
M.Szyd\l{}owski, J.Szcz\c{e}sny, Phys.Rev. D {\bf 50} (1994) 819.\\
\bibitem{Anosov}
D.W.Anosov, Trudy Mat.Inst.W.A.Steklova, (1967) 90. \\ \
\end{thebibliography}
\end{document}